# Microstructure reconstruction using entropic descriptors

BY RYSZARD PIASECKI[*]

*Institute of Physics, University of Opole, Oleska 48, 45-052 Opole, Poland*

A multi-scale approach to the inverse reconstruction of a pattern's microstructure is reported. Instead of a correlation function, a pair of entropic descriptors (EDs) is proposed for stochastic optimization method. The first of them measures a spatial inhomogeneity, for a binary pattern, or compositional one, for a greyscale image. The second one quantifies a spatial or compositional statistical complexity. The EDs reveal structural information that is dissimilar, at least in part, to that given by correlation functions at almost all of discrete length scales. The method is tested on a few digitized binary and greyscale images. In each of the cases, the persuasive reconstruction of the microstructure is found.

## 1. Introduction

'To what extent can the structure of a disordered heterogeneous material be reconstructed using limited but essentially exact structural information about the original system?' – the first sentence of the abstract in Rintoul & Torquato (1997) still remains a vital question of modelling heterogeneous materials. Much research effort has been concentrated on this topic. One of the particularly useful is the simulated annealing (SA) technique, widely discussed in Torquato (2002*a*,*b*) and Jiao *et al.* (2007, 2008). This technique has the advantage that it is developed for problems with many local minima. The simplest SA-reconstruction of a digitized microstructure by Yeong & Torquato (1998*a*) makes use of the two-point correlation function $S_2(r)$ that represents the probability of finding two particles (pixels) of the phase of interest separated by a distance *r*. Another concept based on incorporating the fast Fourier transform algorithm to calculate $S_2$, as developed in Cule & Torquato (1999) and Fullwood *et al.* (2008*a*,*b*), results in very close replicas of regular and more varied two-phase microstructures up to a translation. On the other hand, Kumar *et al.* (2006, p.820) clearly stated: 'the reconstruction process is not meant to exactly duplicate the parent (target) microstructure, which is already at hand, but rather to create statistically similar microstructures…'. Very recently, a genetic algorithm and maximum entropy method has been compared with SA and a *hybrid* approach based on genetic algorithms and SA has been proposed by Patelli & Schuëller (2009).

    The SA proceeds to find a statistically reasonable realization by evolving the microstructure in such a manner that minimizes 'energy' function *E* taken as squared difference between the correlation functions of the target (reference) and trial (generated) patterns. For isotropic media, this is a quite effective approach except the dense systems with significant aggregation of the particles like the sticky-disk model of Rintoul & Torquato (1997) or multi-scale patterns in Jiao *et al.* (2008). Such structurally complex systems show characteristic features at certain length scales. The Jiao *et al.* (2008) $S_2$-reconstruction of a binary (black and white) complex laser-speckle pattern revealed that the $S_2(r)$ alone could not capture in a satisfactory way all its structural features. Owing to the limited information contained in a single correlation function, the reconstructed microstructure is not unique. In general, a multi-scale

---

[*] piaser@uni.opole.pl

structure cannot be fully characterized with a single lower order correlation function; see Yeong & Torquato (1998*a*). The situation becomes more satisfactory when the second correlation function, e.g. the lineal-path function $L(r)$ studied in Lu & Torquato (1992) is also included in a hybrid reconstruction procedure (Yeong & Torquato 1998*a*; Torquato 2002*a*,*b*; Kumar *et al.* 2006; Jiao *et al.* 2007, 2008). By evaluation of the probability of finding an entire line segment of length *r* within the phase of interest, the $L(r)$ gives information such as the lineal clustering or a coarse level of the connectedness of the microstructure. Recently, a new hybrid $\{S_2(r); C_2(r)\}$-reconstruction has been tested successfully on many textures (Jiao *et al.* 2009). The approach incorporates two-point cluster function $C_2(r)$ developed by Torquato *et al.* (1988). This function gives the probability of finding two points separated by a distance *r* in the same cluster of the phase of interest. Thus it can serve as a sensitive structural indicator when clustering and phase connectedness appear.

Motivated by these observations, certain entropic descriptors (EDs; Piasecki 2000*a*,*b*, 2009*a*; Piasecki & Plastino 2010) are employed for the first time to innovative reconstruction of a pattern's microstructure. The EDs are able to detect relatively dissimilar pattern's features (cf. figure 5 in Piasecki 2009*b*) compared with $S_2$-correlation function. The unbiased hybrid reconstruction (UHR) method we propose provides encouraging results not only for binary (0–1) images. The UHR method is with no trouble applicable also to complex greyscale (0–255) images. Thus, a fresh view is created in the context of reconstructing random media for predicting their effective physical properties and performance optimization; see the recent reviews of Wang & Pan (2008), Fullwood *et al.* (2010) and Torquato (2010).

In this paper, I would like to concentrate on the specific and novel application of EDs, i.e. microstructure reconstruction, not on the other possible categories of spatio-compositional inhomogeneity or spatio-compositional complexity. The introduction to those broad topics can be found in the latest articles (Piasecki 2009*b*, Piasecki & Plastino 2010). This is a reason why only the list of the appropriate formulas and important details are given in appendix. The rest of this paper is organized as follows. In §2, the different EDs are briefly introduced. Next, in §3 the basic versions of the UHR method are formulated and described. Then, the method is examined on some examples of surrogate patterns in §4. Section 5 is devoted to final conclusions and suggestions for further development of the hybrid approach.

## 2. Entropic descriptors

In the simplest version, the proposed hybrid reconstruction of a binary pattern needs the ED-pair $\{S_\Delta; C_\lambda(S)\} \equiv \{S_\Delta; C_{\lambda, S}\}$ of an average (per cell) *spatial* inhomogeneity introduced in Piasecki (2000*a*,*b*) and *spatial* statistical complexity considered by Piasecki & Plastino (2010), respectively. The second ED-pair we use is $\{S_{gr, \Delta}; C_\lambda(S_{gr})\} \equiv \{G_\Delta, C_{\lambda, G}\}$ and basically relates to greyscale images. Now, the $G_\Delta$-component quantifies an average grey level inhomogeneity (Piasecki 2009*a*,*b*), the so-called *compositional* inhomogeneity. The $C_{\lambda, G}$-part quantifies an average grey level statistical complexity (Piasecki & Plastino 2010), the so-called *compositional* statistical complexity. Each of the descriptors makes the use of microcanonical entropy $\text{Entr} = k_B \ln \Omega$, where $k_B$ is equal to unity and $\Omega$ denotes the number of microstates realizing a macrostate properly defined (appendix A).

A configurational (binary) macrostate can be simply described with the help of the order dependent set of cell occupation numbers $\{n_i(k)\}$, $i = 1, 2,…, \kappa(k)$, of black pixels (finite size $1 \times 1$-objects) inside *i*th sliding sampling cell of size $k \times k$. The side length of the cell defines the discrete length scale *k*. Here $\kappa(k) = (L - k + 1)^2$ is the length scale depending number of allowed positions of the sliding cell (with maximal overlapping) for a given pattern of size $L \times L$. In turn, the set of *i*th cell sums $\{g_i(k)\}$ of grey level values determines a compositional (grey level) macrostate. Notice that for the latter case, all possible order dependent partitions allowing some of the parts to be zero of $g_i(k)$ over $k^2$ unit cells inside *i*th cell are referred to as a *weak* composition (Stanley 2001); additional details can be found in Piasecki (2009*a*,*b*).

In our approach, for any length scale $1 \leq k \leq L$ the binary $S_\Delta$-component (the grey level counterpart $G_\Delta$) of the ED-pairs takes into account the statistical *dissimilarity* of actual (current) macrostate AM(*S*) (AM(*G*)) and reference (theoretical) one RM$_{max}$(*S*) (RM$_{max}$(*G*)) that maximizes appropriate entropy. Thus, it is natural to consider the difference of the corresponding entropies. The general form for the entropic binary descriptor reads therefore

$$S_\Delta(k) = \frac{[\text{Entr}_{max}(S) - \text{Entr}(S)]}{\kappa}, \tag{2.1}$$

while for the grey level case

$$G_\Delta(k) = \frac{[\text{Entr}_{max}(G) - \text{Entr}(G)]}{\kappa}. \tag{2.2}$$

To simplify the notation, the variable *k* is omitted on the right-hand side of equations (2.1)–(2.4). For a given pattern, the averaging procedure allows one to compare the descriptor values at different length scales *k*. Of course, for a given binary (greyscale) pattern the form of entropy in equations (2.1)–(2.4) should be specified adequately to each case. Here only the main idea of the EDs is presented. The more detailed description can be found in Piasecki (2000*a*,*b*, 2009*a*,*b*). However, to increase an accessibility of the present method, the formulas used for computing of the numbers $\Omega$ of realizations of appropriate macrostates are given in appendix A.

Consecutively, the binary $C_{\lambda, S}$-component of the ED-pairs takes into consideration the statistical *dissimilarity* of macrostates in the pairs: AM(*S*) and RM$_{max}$(*S*), AM(*S*) and RM$_{min}$(*S*), RM$_{max}$(*S*) and RM$_{min}$(*S*), and in the similar way for the grey level counterpart $C_{\lambda, G}$. In particular, we are interested in those structural features that depend on the length scale *k*. This type of entropic descriptor is able to distinguish structurally distinct configurational (compositional) macrostates with identical or nearly the same degree of spatial (compositional) disorder. The general form of the entropic binary descriptor is given by

$$C_{\lambda, S}(k) = \frac{1}{\kappa} \frac{[\text{Entr}_{max}(S) - \text{Entr}(S)][\text{Entr}(S) - \text{Entr}_{min}(S)]}{[\text{Entr}_{max}(S) - \text{Entr}_{min}(S)]}, \tag{2.3}$$

while for the grey level case

$$C_{\lambda, G}(k) = \frac{1}{\kappa} \frac{[\text{Entr}_{max}(G) - \text{Entr}(G)][\text{Entr}(G) - \text{Entr}_{min}(G)]}{[\text{Entr}_{max}(G) - \text{Entr}_{min}(G)]}. \tag{2.4}$$

For any length scale *k* each of the above statistical complexities vanishes for the two opposite extremes: (i) the corresponding maximum inhomogeneity, when Entr $\rightarrow$ Entr$_{min}$, and (ii) the corresponding maximum homogeneity, when Entr $\rightarrow$ Entr$_{max}$. In between these two special instances, the highest value of spatial or compositional statistical complexity exists: $C_{\lambda, max}(\Omega)|_{\Omega = \Omega_0} = (\ln (\Omega_{max} / \Omega_{min})) / 4\kappa$ for $\Omega_0 = (\Omega_{max} \Omega_{min})^{1/2}$ given here in rather enlightening $\Omega$-notation. In pattern's language, the most statistically complex arrangement at a given length scale emerges when the average *departure* of the actual entropy Entr from its maximum possible value Entr$_{max}$ is comparable to that from its minimum possible value Entr$_{min}$ (Piasecki & Plastino 2010). Further details can be found in Piasecki & Plastino (2010).

## 3. The unbiased hybrid reconstruction

Let us denote by $S^0_\Delta(k)$ ($G^0_\Delta(k)$) and $C^0_{\lambda,S}(k)$ ($C^0_{\lambda,G}(k)$) the binary (greyscale) target EDs computed at a given length scale *k*. Their counterparts, i.e. the EDs for trial patterns will be marked by the corresponding symbols without the superscript zero. Within the present approach, the aforementioned energy function can be taken in any of its hybrid forms $E \equiv E_j$, where $j = S$, $G$ and $M$. As before the '*S*' and '*G*' refer respectively to a binary and greyscale image while the '*M*' relates to a binary pattern that is encoded in two ways: (i) the standard one (0 = black, 1 = white) and (ii) the greyscale fashion (0 = black, 255 = white). Instead of two target EDs, the latter 'twofold' encoding incorporates the set of four EDs. This allows for obtaining a higher structural accuracy in the simulation procedure−for a given tolerance $\delta$-value mentioned below−in comparison to that obtained by means of only two target EDs. This technique has been used in the numerical example 4.1.

The $E_j$ can be expressed as the weighted, by the parameter $1/m$, sum of squared differences between target EDs and those computed for trial configurations

$$E_j = \frac{1}{m}\sum_{k=1}^{L}\varepsilon_j(k), \tag{3.1}$$

where $m = 2$ for instances given by equations (3.2) and (3.3), and $m = 4$ for equation (3.4) is the number of different EDs. The length scale depending terms $\varepsilon_j(k)$ in equation (3.1) are specified as

$$\varepsilon_S(k) = [(S_\Delta - S^0_\Delta)^2 + (C_{\lambda,S} - C^0_{\lambda,S})^2], \quad m = 2 \tag{3.2}$$

$$\varepsilon_G(k) = [(G_\Delta - G^0_\Delta)^2 + (C_{\lambda,G} - C^0_{\lambda,G})^2], \quad m = 2 \tag{3.3}$$

and

$$\varepsilon_M(k) = \varepsilon_S(k) + \varepsilon_G(k), \quad m = 4. \tag{3.4}$$

Considering every configuration as a 'state' of the system, 'energy' *E* can be described as a function of the states; cf. Jiao *et al.* (2007) for $S_2$-reconstruction.

Now, to minimize an objective function *E* we repeat the following steps. For a current configuration of binary (greyscale) pattern two randomly selected pixels of different phases (different grey levels) are interchanged giving the new trial state. The new configuration is then accepted with probability $p(\Delta E)$ given by the Metropolis acceptance rule, see its description in Torquato (2002*a*),

$$p(\Delta E) = \begin{cases} 1 & \Delta E \leq 0, \\ \exp(-\Delta E/T), & \Delta E > 0, \end{cases} \tag{3.5}$$

where $\Delta E = E_{\text{new}} - E_{\text{old}}$ is the change in the energy between the two successive states. Upon acceptance, the trial pattern becomes a current one, and this simple evolving procedure is repeated. The variation of a fictitious temperature *T* as a function of time is called the cooling schedule associated with the annealing process. We use the popular cooling schedule $T(l)/T(0) = \gamma^l$ with a positive parameter $\gamma = 0.8$, where *l* numerates annealing steps. Note that if the thermalization (the system should evolve long enough at $T(l)$) and annealing rate (the closer $\gamma$ to one the slower annealing process) are not carefully chosen then one can obtain sub-optimal results. However, for practical test purposes, the above cooling schedule is sufficient (Jiao *et al.* 2007). We terminate the reconstruction when energy *E* becomes smaller than such a tolerance value $\delta$, for which throughout annealing loop *l* at temperature $T(l)$ the proportion of accepted Monte Carlo steps (equation 3.5) to its *l*-depending entire number becomes less than approximately 0.01. Thus, the distinct $\delta$-values appeared in order to avoid a considerable increase in computation time that depends mainly on the pattern's size and also on the type of hybrid approach. It should be stressed that in the later stages of SA no special technique, as discussed in Kumar *et al.* (2006) and Jiao *et al.* (2008) modifying the exchange of isolated pixels or requiring the use of a biased pixel-selection process in either of the phases has been used. Therefore the proposed approach belongs to the *unbiased* annealing technique.

## 4. Illustrative examples

The capability for the inverse reconstruction of various microstructures by the proposed UHR method is examined on exemplary images.

**Example 4.1.** Firstly, the UHR method is examined for a complex binary laser-speckle pattern that belongs to a class of multi-scale patterns. The black phase volume fraction of our target pattern is $\varphi_1 \cong 0.646$; see the inset T #1 in figure 1*b*. This $64 \times 64$ sub-domain is adapted from a $129 \times 129$-version of the target pattern (cf. upper left corner of figure 12*a* in Jiao *et al.* 2008). Making use of the already mentioned two-fold encoding of a binary pattern, the two ED-pairs, $\{S_\Delta, C_{\lambda, S}\}$ and $\{G_\Delta, C_{\lambda, G}\}$, contribute in this instance. The UHR procedure has been terminated when $E \leq \delta = 2 \times 10^{-2}$. It is worth to notice that the total number of evaluations of objective function given by equations (3.1)–(3.4) equals to about $2 \times 10^5$ in this case. In figure 1*a* the solid lines (symbols) correspond to appropriate EDs computed for target pattern T #1 (its reconstruction R #1), respectively. In this example, the fitting is sufficient for a structurally convincing target pattern's reconstruction (see the inset R #1 in figure 1*b*), at least with qualitative visual assessment.

It is instructive to consider now the behaviour of scaled autocovariance function $f(r)$, which is defined as $f(r) = (S_2(r) - \varphi_1^2)/(\varphi_1(1 - \varphi_1))$, where $S_2$ is the two-point correlation function of black phase. Let us make the preliminary comparison for T #1, its UHR-reconstruction R #1 and four non-overlapping $64 \times 64$ sub-domains A($\varphi_1 \cong 0.656$), B($\varphi_1 \cong 0.674$), C($\varphi_1 \cong 0.598$) and D($\varphi_1 \cong 0.630$); see the proper pictures under table 1. The sub-domains are taken from the pattern (cf. figure 13 in Jiao *et al.* 2008) being the $S_2$-reconstruction of the $129 \times 129$ laser-speckle pattern. For simplicity, the function $S_2$ is evaluated along two orthogonal directions (the rows and columns of pixels). This orthogonal-sampling algorithm introduced by Yeong & Torquato (1998*a,b*) is consequently applied in this paper as well as the hard-wall conditions.

Notice, that figure 1*a* shows very well fitting of the target ED-curves with the corresponding symbols for the reconstructed pattern at every length scale *k*. In spite of that, now in figure 1*b* the behaviour of two *f*-counterparts, i.e. thick solid line for T #1 and thick dashed one for R #1 is dissimilar at almost all of discrete length scales. Earlier, the reverse lack of similarity was signalized, i.e. between $S_\Delta$-curves as well as $G_\Delta$-lines (cf. figure 5 in Piasecki 2009*b*) computed for the $S_2$-reconstructed pattern (Jiao *et al.* 2008). Thus, it is clear that limited structural information provided by function $S_2$ and by the EDs is comparatively different. In figure 1*b* also a group of four thin solid lines, which correspond to the A, B, C and D sub-domains, is added for comparison purposes. Each of the lines shows a more or less similar mutual dissimilarities but the reason is a different, at least in part. Namely, when a sub-domain is examined instead of the whole pattern, usually the structural integrity is a bit decreased and thus the *f*-convergence changes for the worse.

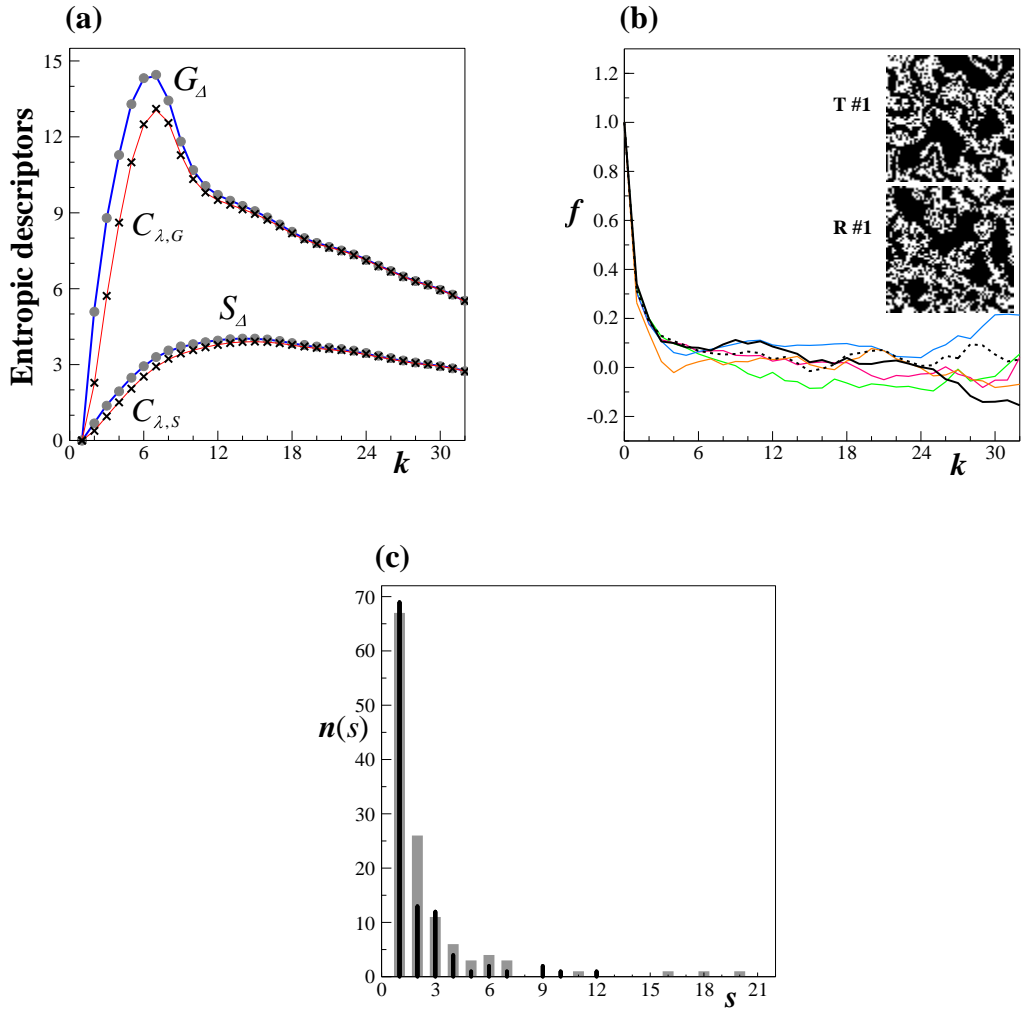

**Figure 1.** (*a*) The unbiased hybrid reconstruction (UHR) making use of two pairs of EDs, $\{S_\Delta; C_{\lambda,S}\}$ and $\{G_\Delta; C_{\lambda,G}\}$, for a $64 \times 64$ sub-domain T #1 of the binary laser-speckle pattern adapted from Jiao *et al.* (2008) (with the permission of the authors). The lines correspond to target pattern T #1 while the symbols refer to its reconstruction R #1. For better visualization of details we limit the length scales to $k \leq 32$. (*b*) The scaled autocovariance function *f* of black phase computed for T #1 (thick solid line) and R #1 (dashed line), and also for the A, B, C and D sub-domains (thin solid lines: green, rose, blue and orange on-line) of the whole laser-speckle pattern reconstructed by Jiao *et al.* (2008). (*c*) The comparison of distribution of cluster sizes *s* (in pixels) for T #1 (black dropped lines) and R #1 (grey ones). The number $n(s)$ of clusters of size $s > 22$ is given in table 1.

It is worth noticing that certain quantitative arguments about the quality of any inverse reconstruction can be extracted from a distribution of cluster sizes (two pixels aligned along the diagonal form two clusters each of size one). At present case, the number $n_1$ of isolated black pixels for the R #1-reconstruction ($n_1 = 67$) is closer to that for the target pattern T #1 ($n_1 = 69$) compared to the four sub-domains: A ($n_1 = 48$), B ($n_1 = 51$), C ($n_1 = 81$) and D ($n_1 = 75$) (table 1). However, the evaluation for clusters of higher sizes is less clear. One can state that multi-scale structural elements like the small black clusters are reconstructed by the present UHR method much better than the black stripes; see figure 1*c*. A similar observation seems to be true also for the $S_2$-reconstruction alone.

**Table 1.** The numbers $n(s)$ of cluster sizes $s$ (in pixels) are presented for the target pattern T #1 and its UHR-reconstruction R #1. Additionally, for preliminary comparison purposes, the number $n(s)$ for the four non-overlapping $64 \times 64$ sub-domains marked here as $\begin{smallmatrix}AB\\CD\end{smallmatrix}$ is given. The exemplary sub-domains are taken from $S_2$-reconstruction (Jiao *et al.* 2008). Each of the patterns is depicted below the right column.

| T #1 | | R #1 | | A | | B | | C | | D | |
|---|---|---|---|---|---|---|---|---|---|---|---|
| *s* | *n* | *s* | *n* | *s* | *n* | *s* | *n* | *s* | *n* | *s* | *n* |
| 1 | 69 | 1 | 67 | 1 | 48 | 1 | 51 | 1 | 81 | 1 | 75 |
| 2 | 13 | 2 | 26 | 2 | 10 | 2 | 2 | 2 | 8 | 2 | 9 |
| 3 | 12 | 3 | 11 | 3 | 6 | 3 | 6 | 3 | 4 | 3 | 3 |
| 4 | 4 | 4 | 6 | 4 | 1 | 4 | 2 | 4 | 3 | 4 | 6 |
| 5 | 1 | 5 | 3 | 5 | 1 | 5 | 1 | 5 | 4 | 5 | 1 |
| 6 | 2 | 6 | 4 | 6 | 2 | 6 | 1 | 6 | 4 | 7 | 2 |
| 7 | 1 | 7 | 3 | 7 | 2 | 7 | 1 | 7 | 3 | 8 | 1 |
| 9 | 2 | 11 | 1 | 8 | 1 | 8 | 1 | 9 | 2 | 12 | 1 |
| 10 | 1 | 16 | 1 | 13 | 1 | 10 | 1 | 10 | 1 | 13 | 1 |
| 12 | 1 | 18 | 1 | 14 | 1 | 11 | 1 | 11 | 1 | 14 | 2 |
| 32 | 1 | 20 | 1 | 23 | 1 | 16 | 1 | 12 | 2 | 18 | 1 |
| 55 | 1 | 38 | 1 | 34 | 1 | 25 | 1 | 17 | 1 | 22 | 1 |
| 58 | 1 | 51 | 1 | 101 | 1 | 2593 | 1 | 18 | 1 | 24 | 1 |
| 60 | 1 | 62 | 1 | 424 | 1 | | | 19 | 1 | 27 | 1 |
| 75 | 1 | 88 | 1 | 1950 | 1 | | | 29 | 1 | 2284 | 1 |
| 2155 | 1 | 257 | 1 | | | | | 33 | 1 | | |
| | | 1849 | 1 | | | | | 2085 | 1 | | |

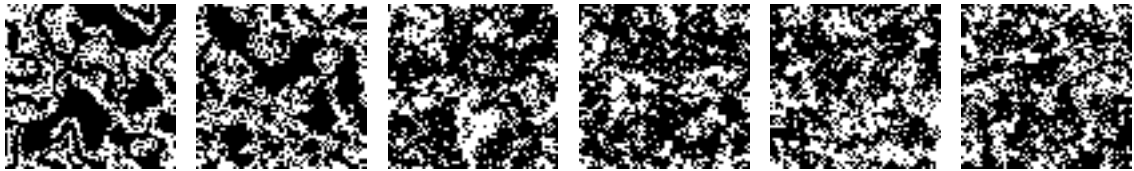

**Example 4.2.** Consider now a modified $42 \times 42$ domain of greyscale image adapted from Noussiou & Provata (2007). There a surface reconstruction in reactive dynamics within mesoscopic kinetic Monte-Carlo approach was performed. For illustrative reasons we convert the domain into a three-level ($255 =$ white, $127 =$ grey and $0 =$ black) target pattern by specifying two greyscale thresholds, which lead to the equal; see the inset T #2 in figure 2*a*. The connotation of each of the colours is irrelevant here, as it has nothing to do with the purposes of the UHR presentation. It is worth noticing that the most compact clusters form black pixels. The UHR procedure begins with the random configuration of black, grey and white pixels. The tolerance value $\delta = 4 \times 10^{-4}$ means the ending of the procedure. Such a tolerance value, notwithstanding the usage of a one pair of compositional EDs, in this case the $\{G_\Delta, C_{\lambda,G}\}$, yields a structurally persuasive target pattern's reconstruction as depicted in the inset R #2 in figure 2*a*. Keeping unchanged distribution of the black phase and interchanging the more spread-out white with the grey phase, a one additional target pattern T ## (not shown here) can be obtained. Its reconstruction shows the additional inset R ## in figure 2*a*. The comparison of values of EDs(T 2#) with EDs(T ##) as well as EDs(R 2#) with EDs(R ##) reveals that the subtle structural differences between arrangements of white and grey pixels lead to a higher compositional inhomogeneity and statistical complexity of the additional patterns; see http://arxiv.org/abs/0910.1955v4 (cond-mat.stat-mech). The detection of such effect by the naked eye inspection of the patterns seems to be impossible. Thus, the UHR approach employing merely one pair of EDs is still sensitive tool even for multi-phase media.

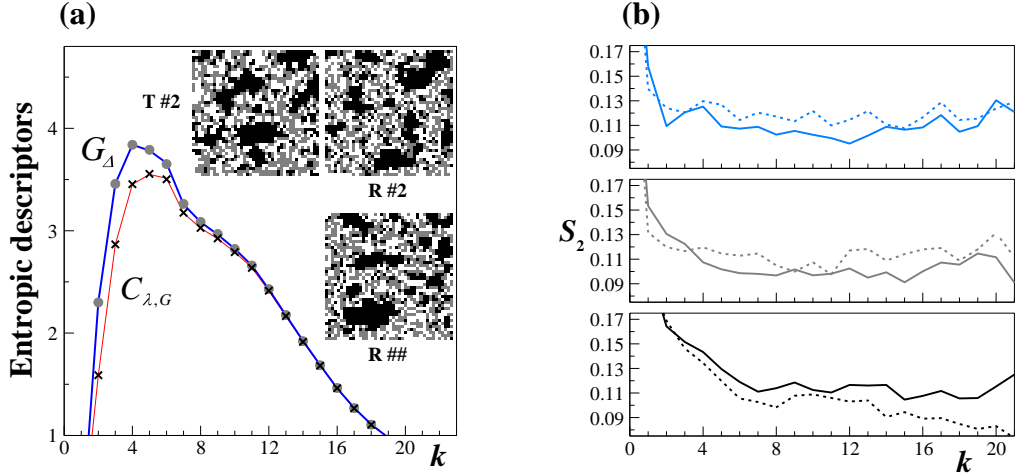

**Figure 2.** (*a*) The UHR making use of a pair of EDs, {$G_\Delta$; $C_{\lambda,G}$}, for a modified $42 \times 42$ domain of three-level image presented in Noussiou & Provata (2007). The lines correspond to target pattern T #2 while the symbols refer to its reconstruction R #2. For better visualisation of details we show only the EDs values larger than one and limit the length scales to $k \leq 23$. (*b*) From top to bottom, the $S_2$-function (within interesting range of its values) of a white, grey and black phase computed for T #2 (solid lines) and R #2 (dashed lines).

In figure 2*b*, the quality of the UHR reconstruction is enlightened from a different viewpoint. Since the volume fractions of white, grey and black phases are identical there is no need to use scaled autocovariance function *f*. Therefore, the pairs of {$S_2$(T 2#) (solid lines); $S_2$(R 2#) (dashed lines)} for a white (top), grey (middle) and black (bottom) phase are presented within the interesting range of its values. The $S_2$ convergence is not satisfactory *both* for the target pattern and its reconstructions. This indicates the need for consideration of patterns larger in size. In this way the higher structural integrity of the target and reconstructed patterns should be ensured. Similar to the previous example, for each of the pairs the lack of similarity appears over nearly all of length scales. So, also for multi-phase media we came to the same as for binary patterns conclusion that the correlation function $S_2$ and EDs provide relatively dissimilar structural information.

**Example 4.3.** Now focus on a fully grey level pattern. Avoiding mathematical details, which are not a subject of the present work, a greyscale pattern without a clear symmetry (see references for the link to an accompanying animation) is adapted from Rucklidge & Silber (2009). The pattern is a transient one between the 12-fold and 14-fold approximate quasi-patterns. The authors have investigated the time-dependent model partial differential equation involving the pattern-forming field $U(x, y, t)$ being a complex-valued function and real-valued $2\pi$-periodic forcing function $f(t)$; cf. equation (3.1) in Rucklidge & Silber (2009). In resulting patterns the greyscale represents the real part of $U(x, y, t)$. In order to reduce the computation cost, the $137 \times 137$ sub-domain exemplary for the transient case was resized to $69 \times 69$. This modified greyscale pattern is chosen as the target one; see the middle inset T #3 in figure 3*b*. The UHR procedure uses a one pair of compositional EDs, {$G_\Delta$, $C_{\lambda,G}$}. As usual, it starts with an initial random configuration. Now, this is a pattern I #3 (the top inset in figure 3*b*) of the same grey level histogram as the T #3. The reconstruction has been terminated when $E \leq \delta = 3 \times 10^{-5}$.

In figure 3*a*, the solid $G_\Delta$-line and symbols refer to target pattern T #3 and its reconstruction R #3; see the bottom inset in figure 3*b*. For comparison, the dashed line, which corresponds to initial pattern I #3 is also shown. The inset in figure 3*a* shows that this time the $C_{\lambda,G}$-line is hardly distinguishable from the $G_\Delta$-one even around the first peak. Therefore, it is absent on the main part of the figure. Similarly to the earlier examples, one can observe that the UHR method applied to entirely greyscale pattern still permits for structurally credible its reconstruction, at least based upon qualitative visual assessments of greyscale structure for T #3 and R #3. The high structural similarity appears to be retrieved at *every* scale in a strikingly convincing way (figure 3*b*). This supports our belief that the present approach is quite universal one although relatively time consuming. Increasing the efficiency of the program code the computational time could be considerably reduced. However, to obtain statistically significant number of reconstructions, especially for patterns larger in size, it is crucial the use of a more powerful computer than a personal one.

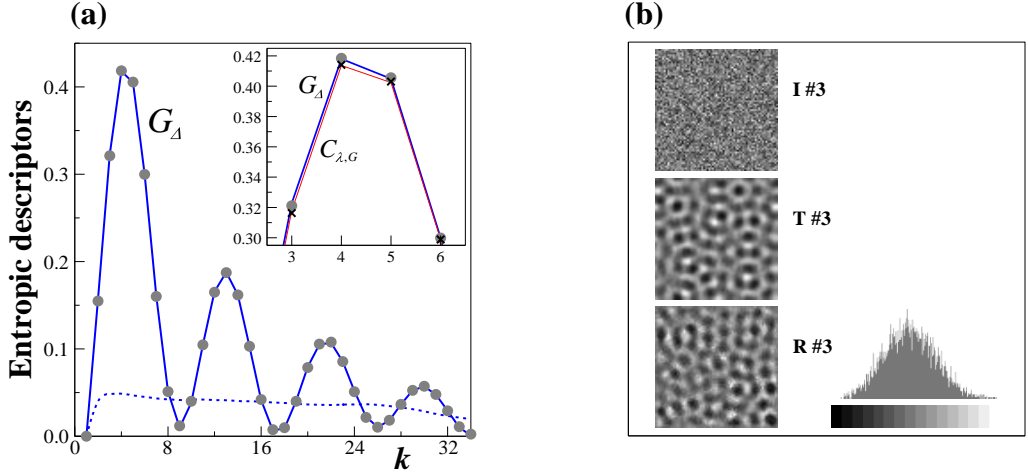

**Figure 3.** (*a*) The UHR making use of a pair of EDs, $\{G_\Delta; C_{\lambda, G}\}$, for an exemplary $137 \times 137$ sub-domain (resized then to $64 \times 64$) of greyscale intermediate pattern adapted from the animated movie of the transition between 12-fold and 14-fold approximated quasi-patterns investigated in Rucklidge & Silber (2009) (with the permission of the authors). The solid $G_\Delta$-lines correspond to target pattern T #3, the symbols refer to its reconstruction R #3 while the dashed line applies to the initial random pattern I #3. In the inset, the $C_{\lambda, G}$-line hardly distinguishable from $G_\Delta$-one is also shown around the first peak. (*b*) To facilitate comparison of the patterns, the initial one I #3 (top), target T #3 (middle) and its reconstruction R #3 (bottom) are depicted. In addition, the common grey level histogram is also shown.

## 5. Conclusions

We gather that the UHR is a generally applicable method. In particular, the results within this approach are readily available for any random binary media as well as for grey level patterns. Other microstructure reconstructions for a variety of patterns, not yet presented here, support this observation. The main conclusions to be drawn can be summarized as follows: in the vast majority of cases (excluding possible pathological), the microstructural information revealed by means of the EDs allows for innovative inverse reconstruction with nearly all multi-scale structural elements recaptured on an acceptable level in connection with a given tolerance $\delta$-value. Our opinion about the quality of the reconstruction with the new method, preferably at small length scales, is based mainly on qualitative visual assessments of structure. It is also supported in part by quantitative evaluations like cluster sizes distributions given in table 1 and drawn in figure 1*c*, behaviour at initial length scales of orthogonal autocovariance function *f* shown in figure 1*b* and correlation function $S_2$ in figure 2*b*. On the other hand, for an entirely greyscale case, the strikingly convincing structural similarity between the target quasi-pattern and its reconstruction seems to appear at *every* scale; see the proper insets in figure 3*b*. In general, for the inverse reconstruction methods based on SA but making use of different objective functions, one can expect a higher structural 'accuracy' at wider range of length scales for patterns larger in size. It is also worth noticing that our method can be extended to *q*-entropies (Piasecki *et al.* 2002) that are being the subject of much work in statistical mechanics (Tsallis 2009).

Finally, one can take for granted that two-point correlation function and EDs provide relatively dissimilar structural information. Thus, especially for random binary media, a simple way of further improving of the reconstruction process can be considered. For instance, a pair of the EDs can be combined with a pair of the two-point correlation and cluster functions. Potentially useful complementary structural information could be revealed effectively by this sort of a doubly hybrid approach.

## Acknowledgements

I would like to thank Angelo Plastino for careful reading an earlier version of this paper and Wiesław Olchawa for providing clusters distribution procedure.

## Appendix A

In general, given binary (S) or grey level (G) pattern of size $L \times L$ can be sampled by $\kappa(k) = ((L-k)/z + 1)^2$ cells of size $k \times k$ with a sliding factor $1 \leq z \leq k$ provided $((L-k) \bmod z) = 0$. Here the $z = 1$ is chosen that gives the maximal overlapping of the cells. In fact, in this way we analyse auxiliary patterns $L_a(k) \times L_a(k)$, where $L_a(k) \equiv ((L-k)/z + 1)\,k$. Those patterns composed of the sampled cells placed in a non-overlapping manner can be treated as the representative ones since they clearly reproduce the general structure of the initial images; see http://arxiv.org/abs/0910.1955v4 (cond-mat.stat-mech). Such approach allows us to compute the reference entropies, $\text{Entr}_{\max}$ and $\text{Entr}_{\min}$, which are related to specific macrostates of the representative patterns. Keeping this in mind, the basic constraints at every length scale $k$ for cell occupation numbers $n_i(k)$ and for local grey level sums $g_i(k)$ can be written as:

$$\text{(i)} \quad \sum_{i=1}^{\kappa} n_i(k) = N(k), \quad \text{and} \quad \text{(ii)} \quad \sum_{i=1}^{\kappa} g_i(k) = G(k), \tag{A 1}$$

where $N(k)$ and $G(k)$ stand for the total number of black pixels and the total sum of grey level values, both dependent on length scale. We assume that the black pixels concentration $\varphi_S = N(k)/\kappa k^2$ as well as the normalized average (per cell) sum of grey level values $\varphi_G = G(k)/255\kappa k^2$ are non-trivial. That means the following inequalities are obeyed: $0 < \varphi_S < 1$ and $0 < \varphi_G < 1$. To simplify notation we will omit the parameter $k$ wherever it does not leads to misunderstanding.

At every fixed length scale $k$, only the numbers of realizations of appropriate macrostates are listed below. The first three of them, i.e. equations (A 2)–(A 4), refer to binary patterns and the remaining ones, i.e. equations (A 5)–(A 7), relate to grey level case. We begin with the clear number $\Omega(S)$ of realizations of the actual macrostate AM(S) that is the product of the ways that each of sampled cells composed of $k^2$ unit cells can be occupied with the number $n_i$ of black pixels under above constraint (i)

$$\Omega(S) = \prod_{i=1}^{\kappa} \binom{k^2}{n_i}. \tag{A 2}$$

The maximum possible value $\text{Entr}_{\max}(S)$ is accessible for most spatially homogeneous reference macrostate, $\text{RM}_{\max}(S) \equiv \{n_i \in (n_0, n_0 + 1)\}_{\max}$, with $\kappa - r_0$ and $r_0$ number of cells occupied by $n_0 \in (0, 1, ..., k^2 - 1)$ and $n_0 + 1$ of black pixels. Thus, the simple relation holds: $N = (\kappa - r_0)\{n_0\} + r_0\{n_0 + 1\} \equiv \kappa n_0 + r_0$, where $r_0 = (N \bmod \kappa)$, $r_0 \in (0, 1, ..., \kappa - 1)$ and $n_0 = (N - r_0)/\kappa$. The number of proper microstates then reads

$$\Omega_{\max}(S) = \binom{k^2}{n_0}^{\kappa - r_0} \binom{k^2}{n_0 + 1}^{r_0}. \tag{A 3}$$

In turn, the minimum possible value $\text{Entr}_{\min}(S)$ is available for most spatially inhomogeneous reference macrostate, $\text{RM}_{\min}(S) \equiv \{n_i \in (0, n, k^2)\}_{\min}$, with $\kappa - q_0 - 1$ of empty cells, one cell with $n_i = n \in (0, 1, ..., k^2 - 1)$ and $q_0$ of fully occupied cells. Now, another relation holds: $N = (\kappa - q_0 - 1)\{0\} + 1\{n\} + q_0\{k^2\} \equiv n + q_0 k^2$, where $n = (N \bmod k^2)$, and $q_0 = (N - n)/k^2$, $q_0 \in (0, 1, ..., \kappa - 1)$. The number of proper microstates is therefore

$$\Omega_{\min}(S) = \binom{k^2}{0}^{\kappa - q_0 - 1} \binom{k^2}{n} \binom{k^2}{k^2}^{q_0} \equiv \binom{k^2}{n}. \tag{A 4}$$

Now, we focus on the number $\Omega(G)$ of realizations of the actual macrostate $\text{AM}(G)$. The number of the appropriate compositional microstates is the product of the ways that each of sampled cells can be populated with the number $k^2$ of grey levels under above constraint (ii). Hence, the number of so-called weak compositions (Stanley 2001) is given by

$$\Omega(G) = \prod_{i=1}^{\kappa} \binom{g_i + k^2 - 1}{k^2 - 1}, \tag{A 5}$$

where $g_i$ denotes $i$th cell sum of grey level values.

The maximum possible value $\text{Entr}_{\max}(G)$ relates to most compositionally homogeneous reference macrostate, the $\text{RM}(G) = \{g_i \in (g_0, g_0 + 1)\}_{\max}$, with $\kappa - R_0$ and $R_0$ number of cells having local sums $g_0 \in (0, 1, ..., 255k^2 - 1)$ and $g_0 + 1$ of grey levels. This leads to the simple expression: $G = (\kappa - R_0)\{g_0\} + R_0\{g_0 + 1\} \equiv \kappa g_0 + R_0$, where $R_0 = (G \bmod \kappa)$, $R_0 \in (0, 1, ..., \kappa - 1)$ and $g_0 = (G - R_0)/\kappa$. Thus, the number of the proper microstates reads

$$\Omega_{\max}(G) = \binom{g_0 + k^2 - 1}{k^2 - 1}^{\kappa - R_0} \binom{g_0 + k^2}{k^2 - 1}^{R_0}. \tag{A 6}$$

Finally, the minimum possible value $\text{Entr}_{\min}(G)$ is obtainable for most compositionally inhomogeneous reference macrostate, $\text{RM}_{\min}(G) \equiv \{g_i \in (0, g, 255k^2)\}_{\min}$, with $\kappa - Q_0 - 1$ cells having grey level sums equal to zero, one cell with $g_i = g \in (0, 1, ..., 255k^2 - 1)$ and $Q_0$ cells with $g_i = 255k^2$. Then one can simply write: $G = (\kappa - Q_0 - 1)\{0\} + 1\{g\} + Q_0\{255k^2\} \equiv g + 255 Q_0 k^2$, where $g = (G \bmod 255k^2)$ and $Q_0 = (G - g)/255k^2$, $Q_0 \in (0, 1, ..., \kappa - 1)$. Hence, the number of proper microstates is described by

$$\Omega_{\min}(G) = \binom{0 + k^2 - 1}{k^2 - 1}^{\kappa - Q_0 - 1} \binom{g + k^2 - 1}{k^2 - 1} \binom{255k^2 + k^2 - 1}{k^2 - 1}^{Q_0}. \tag{A 7}$$

The above formulas contain the whole information we need to obtain the corresponding entropy. In order to ease preparing of individual algorithm, I recommend a mathematical identity suitable to compute the logarithm of binomial coefficient; see Van Siclen (1997).

To demonstrate the meaning of the notation a simple binary pattern **P** ⇒ of size $L = 4$ is analysed by a sliding sampling cell $2 \times 2$ at scale $k = 2$. Thus auxiliary pattern $\mathbf{P}(k=2) \equiv \mathbf{P}2$ is composed of $\kappa = 9$ sampled cells and its size equals to $L_a = ((4-2)/1 + 1)\,2 = 6$.

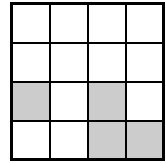

For actual macrostate $AM(S) = \{000111123\}$ of auxiliary pattern **P**2 ⇒

$$\Omega(S) = \binom{4}{0}^3 \binom{4}{1}^4 \binom{4}{2}\binom{4}{3} = 6144, \quad \text{and} \quad Entr(S) = 8.7232$$

For most homogeneous reference macrostate, $RM_{max}(S) = \{111111111\}$

$$\Omega_{max}(S) = \binom{4}{1}^9 = 262144, \quad \text{and} \quad Entr_{max}(S) = 12.4766$$

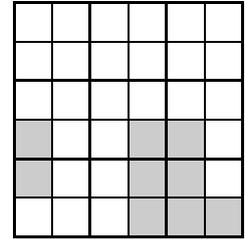

For most inhomogeneous reference macrostate, $RM_{min}(S) = \{000000144\}$

$$\Omega_{min}(S) = \binom{4}{0}^6 \binom{4}{1}\binom{4}{4}^2 = 4, \quad \text{and} \quad Entr_{min}(S) = 1.3863$$

The corresponding values of parameters are: $r_0 = (9 \bmod 9) = 0$, $n_0 = (9-0)/9 = 1$, $n = (9 \bmod 4) = 1$ and $q_0 = (9-1)/4 = 2$.

# References

Cule, D. & Torquato, S. 1999 Generating random media from limited microstructural information via stochastic optimization. *J. Appl. Phys.* **86**, 3428–3437. (doi:10.1063/1.371225)

Fullwood, D. T., Kalidindi, S. R., Niezgoda, S. R., Fast, A. & Hampson, N. 2008a Gradient-based microstructure reconstructions from distributions using fast Fourier transforms. *Mater. Sci. Engin. A* **494**, 68–72. (doi:10.1016/j.msea.2007.10.087)

Fullwood, D. T., Niezgoda, S. R. & Kalidindi, S. R. 2008b Microstructure reconstructions from 2-point statistics using phase-recovery algorithms. *Act. Mater.* **56**, 942–948. (doi:10.1016/j.actamat.2007.10.044)

Fullwood, D. T., Niezgoda, S. R., Adams, B. L. & Kalidindi, S. R. 2010 Microstructure sensitive design for performance optimization. *Prog. Mater. Sci.* **55**, 477–562 (doi: 10.1016/j.pmatsci.2009.08.002)

Jiao, Y., Stillinger, F. H. & Torquato, S. 2007 Modeling heterogeneous materials via two-point correlation functions: Basic principles. *Phys. Rev. E* **76**, (the part I) 031110. (doi:10.1103/PhysRevE.76.031110)

Jiao, Y., Stillinger, F. H. & Torquato, S. 2008 Modeling heterogeneous materials via two-point correlation functions: II. Algorithmic details and applications. *Phys. Rev. E* **77**, (the part II) 031135. (doi:10.1103/PhysRevE.77.031135)

Jiao, Y., Stillinger, F. H. & Torquato, S. 2009 A superior descriptor of random textures and its predictive capacity. *Proc. Natl Acad. Sci. USA* **106**, 17634–17639. (doi:10.1073/pnas.0905919106)

Kumar, H., Briant, C. L. & Curtin, W. A. 2006 Using microstructure reconstruction to model mechanical behavior in complex microstructures. *Mech. Mater.* **38**, 818–832. (doi:10.1016/j.mechmat.2005.06.030)

Lu, B. & Torquato, S. 1992 Lineal path function for random heterogeneous materials. *Phys. Rev. A* **45**, 922–929. (doi:10.1103/PhysRevA.45.922)

Noussiou, W. K. & Provata, A. 2007 Surface reconstruction in reactive dynamics: A kinetic Monte Carlo approach. *Surf. Sci.* **601**, 2941–2951. (doi:10.1016/j.susc.2007.04.258)

Patelli, E. & Schuëller, G. 2009 On optimization techniques to reconstruct microstructures of random heterogeneous media. *Comput. Mater. Sci.* **45**, 536–549. (doi:10.1016/j.commatsci.2008.11.019)

Piasecki, R. 2000a Entropic measure of spatial disorder for systems of finite-sized objects. *Physica A* **277**, 157–173. (doi:10.1016/S0378-4371(99)00458-6)

Piasecki, R. 2000b Detecting self-similarity in surface microstructures. *Surf. Sci.* **454–456**, 1058–1062. (doi:10.1016/S0039-6028(00)00166-7)

Piasecki, R. 2009a A versatile entropic measure of grey level inhomogeneity. *Physica A* **388**, 2403–2709. (doi:10.1016/j.physa.2009.02.031)

Piasecki, R. 2009b Statistical mechanics characterization of spatio-compositional inhomogeneity. *Physica A* **388**, 4229–4240. (doi:10.1016/j.physa.2009.06.028)

Piasecki, R. & Plastino, A. 2010 Entropic descriptor of a complex behaviour. *Physica A* **389**, 397–407. (doi:10.1016/j.physa.2009.10.013)

Piasecki, R., Martin, M. T. & Plastino, A. 2002 Inhomogeneity and complexity measures for spatial patterns. *Physica A* **307**, 157–171. (doi:10.1016/S0378-4371(01)00591-X)

Rintoul, M. D. & Torquato, S. 1997 Reconstruction of the structure of dispersions. *J. Colloid Surface Sci.* **186**, 467–476. (doi:10.1006/jcis.1996.4675)

Rucklidge, A. M. & M. Silber, M. 2009 Design of parametrically forced patterns and quasipatterns. *SIAM J. Appl. Dyn. Syst.* **8**, 298–347. See http://www.maths.leeds.ac.uk/~alastair/papers/RS_qp_siads_example_12_to_14_anim.gif. (doi:10.1137/080719066)

Stanley, R. P. 2001 *Enumerative combinatorics*, vol. I. Cambridge, UK: Cambridge University Press.

Torquato, S. 2002a *Random heterogeneous materials: microstructure and macroscopic properties*. Berlin, Germany: Springer.

Torquato, S. 2002b Statistical description of microstructures. *Annu. Rev. Mater. Res.* **32**, 77–111. (doi:10.1146/annurev.matsci.32.110101.155324)

Torquato, S. 2010 Optimal design of heterogeneous materials. *Annu. Rev. Mater. Res.* **40**, 101–129. (doi:10.1146/annurev-matsci-070909-104517)

Torquato, S., Beasley, J. D. & Chiew, Y. C. 1988 Two-point cluster function for continuum percolation. *J. Chem. Phys.* **88**, 6540-6547. (doi:10.1063/1.454440)

Tsallis, C. 2009 *Introduction to nonextensive statistical mechanics*. Berlin, Germany: Springer.

Van Siclen, C. DeW. 1997 Information entropy of complex structures. *Phys. Rev. E* **56**, 5211–5215. (doi:10.1103/PhysRevE.56.5211)

Wang, M. & Pan, N. 2008 Predictions of effective physical properties of complex multiphase materials. *Mater. Sci. Eng. R.* **63**, 1–30. (doi:10.1016/j.mser.2008.07.001)

Yeong, C. L. Y. & Torquato, S. 1998a Reconstructing random media. *Phys. Rev. E* **57**, 495–506. (doi:10.1103/PhysRevE.57.495)

Yeong, C. L. Y. & Torquato, S. 1998b Reconstructing random media. II. Three-dimensional media from two-dimensional cuts. *Phys. Rev. E* **58**, 224–233. (doi:10.1103/PhysRevE.58.224)